\documentclass[10pt,journal,letterpaper,compsoc]{IEEEtran}
\usepackage{cite}
\usepackage[dvips]{graphicx}
\graphicspath{{./eps/}}
\DeclareGraphicsExtensions{.eps}
\usepackage{amsmath,amssymb,color}
\usepackage{algorithm}
\usepackage{algorithmic}
\usepackage{array}
\usepackage{mdwmath}
\usepackage{mdwtab}
\usepackage{eqparbox}
\usepackage{multirow}

\usepackage[tight,footnotesize]{subfigure}
\usepackage[obeyspaces,spaces]{url}
\usepackage{balance}
\usepackage{hyphenat}
\usepackage{wrapfig}
\begin{document}
%
%================================================================%
%							                TITLE
%================================================================%
%
\title
  {
    IFCIoT: Integrated Fog Cloud IoT Architectural Paradigm for Future Internet of Things
  }
%
%================================================================%
%							                AUTHORS
%================================================================%
%
\author
  {
		\IEEEauthorblockN
    {
    Arslan Munir\IEEEauthorrefmark{1},  %
    Prasanna Kansakar\IEEEauthorrefmark{2} and
    Samee~U.~Khan\IEEEauthorrefmark{3}
    }

    \IEEEauthorblockA
    {
    Email:                                 %
    \IEEEauthorrefmark{1}arslan@unr.edu,
    \IEEEauthorrefmark{2}pkansakar@nevada.unr.edu,
    \IEEEauthorrefmark{3}samee.khan@ndsu.edu
    }
  }
%
%================================================================%
%                            ABSTRACT
%================================================================%
%
\IEEEcompsoctitleabstractindextext{
\begin{abstract}
We propose a novel integrated fog cloud IoT (IFCIoT) architectural paradigm that promises increased performance, energy efficiency, reduced latency, quicker response time, scalability, and better localized accuracy for future IoT applications. The fog nodes (e.g., edge servers, smart routers, base stations) receive computation offloading requests and sensed data from various IoT devices. To enhance performance, energy efficiency, and real-time responsiveness of applications, we propose a reconfigurable and layered fog node (edge server) architecture that analyzes the applications' characteristics and reconfigure the architectural resources to better meet the peak workload demands. The layers of the proposed fog node architecture include application layer, analytics layer, virtualization layer, reconfiguration layer, and hardware layer. The layered architecture facilitates abstraction and implementation for fog computing paradigm that is distributed in nature and where multiple vendors (e.g., applications, services, data and content providers) are involved. We also elaborate the potential applications of IFCIoT architecture, such as smart cities, intelligent transportation systems, localized weather maps and environmental monitoring, and real-time agricultural data analytics and control.
\end{abstract}
\begin{IEEEkeywords}
  Fog computing, edge computing, Internet of things, reconfigurable architecture, radio access network
\end{IEEEkeywords}
}
\maketitle
\IEEEdisplaynotcompsoctitleabstractindextext
\IEEEpeerreviewmaketitle
%
%================================================================%
%                   S. INTRODUCTION AND MOTIVATION
%================================================================%
%
\sloppy
\nohyphens
{
\section{Introduction and Motivation}
\label{section_introduction_and_motivation}
\IEEEPARstart{T}{he} \emph{Internet of things} (IoT) is a network of physical \emph{things}, \emph{objects} or \emph{devices}, such as radio-frequency identification (RFID) tags, sensors, actuators, mobile phones, and laptops. The IoT enables objects to be sensed and controlled remotely across existing network infrastructure, including the Internet; thereby creating opportunities for more direct integration of the physical world into the cyber world. The IoT becomes an instance of cyber-physical systems (CPS) with the incorporation of sensors and actuators in IoT devices. Objects in IoT can possibly be grouped into geographical or logical \emph{clusters}. Various IoT clusters generate huge amounts of data from diverse locations, which advocates the need to process this data more efficiently. Efficient processing of this data can involve a combination of different computation models, such as in situ processing and offloading to surrogate devices and cloud data centers.

\emph{Cloud computing} is an Internet-based computing paradigm that provides ubiquitous and on-demand access to a shared pool of configurable resources (e.g., processors, storage, services, and applications) to other computers or devices. Although cloud computing paradigm is able to handle huge amounts of data from IoT clusters, the transfer of enormous data to and from cloud computers presents a challenge due to limited bandwidth. Consequently, there is a need to process data near data source, and fog computing provides a promising solution to this problem.

\emph{Fog computing} is a novel trend in computing that aims to process data near data source. Fog computing pushes applications, services, data, computing power, and decision making away from the centralized nodes to the logical extremes of a network. Fog computing significantly decreases the data volume that must be moved between end devices and cloud. Fog computing enables data analytics and knowledge generation to occur at the data source. Furthermore, the dense geographic distribution of fog helps to attain better localized accuracy for many applications as compared to the cloud.

Although fog computing alleviates some of the issues facing the realization of future IoT/CPS applications, the \emph{fog nodes} (e.g., edge servers, smart routers, base stations) may not be able to meet performance, throughput, energy, and latency constraints of future IoT/CPS applications unless fog computing architecture is adapted to satisfy these application requirements. This adaptation is needed at both the system-level and the node-level for fog computing. This article aims to address the architectural challenges associated with the realization of scalable IoT and CPS applications leveraging fog computing. The main contributions of this article are as follows:

\begin{itemize}
    \item A novel integrated fog cloud IoT (IFCIoT) architectural paradigm that harnesses the benefits of IoT, fog, and cloud computing in a unified archetype. The IFCIoT architecture promises increased performance, energy efficiency, quicker response time, scalability, and better localized accuracy for future IoT and CPS applications.
    \item We propose an energy-efficient reconfigurable layered fog node (edge server) architecture that will adapt according to fog computing application requirements. The layers of the proposed architecture include application layer, analytics layer, virtualization layer, reconfiguration layer, and hardware layer. The layered architecture facilitates abstraction and implementation for fog computing paradigm that is distributed in nature and where different service, application, data and content providers are involved.
    \item We discuss the potential applications of the IFCIoT architecture, such as smart cities, intelligent transportation systems (ITS), localized weather maps and environmental monitoring, and real-time agricultural data analytics and control.
\end{itemize}

The remainder of this article is organized as follows. Section~\ref{section_distinction_between_cloud_fog_and_edge_computing} elucidates the distinction between cloud, fog, and edge computing. Section~\ref{section_related_work} provides a summary of related work. Section~\ref{section_fog_computing_for_ITS} describes how fog computing can be used for implementation of intelligent transportation systems (ITS). Section~\ref{section_ifciot_architectural_paradigm} presents our proposed IFCIoT architectural paradigm. The fog node architecture for the IFCIoT architectural paradigm is presented in Section~\ref{section_fog_architecture}. Section~\ref{section_reconfigurable_and_adaptive_fog_node_edge_server_architecture_applied_to_its} describes how our proposed fog node architecture can be applied for implementation of ITS. Section~\ref{section_insights_into_other_applications} discusses insights into potential consumer electronics applications. Finally, Section~\ref{section_conclusions} concludes our article.
%
%==========================================================================================%
%                   S. DISTINCTION BETWEEN CLOUD, FOG AND EDGE COMPUTING
%==========================================================================================%
%
\vspace{-2mm}
\section{Distinction between Cloud, Fog, and Edge Computing}
\label{section_distinction_between_cloud_fog_and_edge_computing}
The distinction between cloud, fog, and edge computing has not been elucidated in many relevant scholarly works to the best of our knowledge. To provide readers with a clear understanding of fog computing, we elucidate the distinction between cloud, fog, and edge computing in this section.

\noindent\textbf{Defining Fog Computing:} Fog computing has been defined in a variety of ways in literature by academia and industry. The term fog computing is often associated with Cisco, that is, ``Cisco Fog Computing'' \cite{CiscoFogComp2016}, however, fog computing is open to the community at large. A coalition of industry and academia has founded the ``OpenFog Consortium'' in November 2015 to promote and accelerate adoption of open fog computing \cite{OpenFogConsort2016}. The coalition founders include ARM, Cisco, Dell, Intel, Microsoft and Princeton University. The OpenFog Consortium \cite{OpenFogConsort2016} defines fog computing as: ``Fog computing is a system-level horizontal architecture that distributes resources and services of computing, storage, control and networking anywhere along the continuum from Cloud to Things''.

Yi et al. \cite{YiFCPAHotWeb15} have defined fog computing as: ``Fog computing is a geographically distributed computing architecture with a resource pool consisting of one or more ubiquitously connected heterogeneous devices (including edge devices) at the edge of network and not exclusively seamlessly backed by cloud services, to collaboratively provide elastic computation, storage and communication (and many other new services and tasks) in isolated environments to a large scale of clients in proximity.'' Aazam et al. \cite{AazamFCIoTIEEEPotentials2016} have defined fog computing as: ``Fog computing refers to bringing networking resources near the underlying networks. It is a network between the underlying network(s) and the cloud(s). Fog computing extends the traditional cloud computing paradigm to the edge of the network, enabling the creation of refined and better applications or services. Fog is an edge computing and micro data center (MDC) paradigm for IoTs and wireless sensor networks (WSNs).''

\noindent\textbf{Distinction Between Fog and Cloud Computing:} The word ``fog'' in fog computing conveys the idea of bringing the advantages of cloud closer to the data source (cf. meteorology: fog is simply a cloud that is close to the ground). Cloud computing is usually a model for enabling convenient and on-demand network use of a shared pool of configurable computing resources, such as networks, servers, storage, applications, and services, that may be rapidly provisioned and released with minimal management effort or vendor interaction. Cloud computing permits options for renting of storage and computing infrastructures, business processes, and overall applications. Fog computing extends cloud computing and services to the edge of the network.

Fog computing can be distinguished from cloud computing based on various metrics as discussed in the following \cite{AbdelshkourCisco2015}. The \emph{proximity} of the fog to the end user is one of the main characteristics that differentiates fog from cloud, that is, fog resides at the edge of the network whereas cloud is located within the Internet. Cloud has a centralized \emph{geographical distribution} whereas fog can have a localized or distributed geographical distribution. Cloud computing systems typically consists of only a few resourceful \emph{server nodes} whereas fog comprises of a large number of relatively less resourceful fog nodes. Furthermore, the processing at fog nodes frees up the core network bandwidth, which helps to improve the overall \emph{network efficiency}. The \emph{distance between client and server nodes} in cloud is typically multiple hops whereas clients can connect to fog nodes usually through a single hop. Consequently, fog computing reduces the \emph{latency} of data transmission from IoT devices to the offloaded server because of the proximity of the fog to the end devices as compared to the cloud. Furthermore, cloud computing platforms typically engender higher \emph{delay jitter} for applications as compared to the applications running on fog nodes. Hence, fog computing is more suitable for real-time IoT and CPS applications as compared to cloud computing.

The fog's ability to provide \emph{location-based customization} of content, services, and applications to the IoT devices is another distinguishing characteristic of fog. Cloud, on the contrary, in most cases is unable to deliver specialized content, services, and applications to devices. The location-based customization of services and information is imperative as the information may be relevant in a local context (i.e., proximity of specific geographic coordinates) and may be irrelevant beyond the physical proximity to that location. Finally, cloud provides limited \emph{mobility support} to end devices whereas mobility of end devices is better supported in fog.

Although fog and cloud computing paradigms have clear distinctions, these paradigms are not a replacement for each other. In fact, the fog and the cloud are interdependent and mutually beneficial since certain functions are naturally more advantageous to carry out in the fog while others are better suited to the cloud. The segmentation of what tasks go to the fog and what tasks go to the backend cloud is application specific, and can change dynamically based upon the state of the network, such as processor loads, link bandwidths, storage capacities, fault events, and security threats \cite{OpenFogArchWhitePaper2016}. The cloud provides various services, such as Infrastructure as a service (IaaS), platform as a service (PaaS), and software as a service (SaaS), for organizations that require elastic scale. Fog computing can provide fog as a service (FaaS) to address various business challenges. FaaS may provide services, such as network acceleration, network functions virtualization (NFV), software-defined networking (SDN), content delivery, device management, complex event processing, video encoding, protocol bridging, traffic offloading, cryptography, and analytics platform, \cite{OpenFogArchWhitePaper2016} etc.

\noindent\textbf{Distinction Between Fog and Edge Computing:} The distinction between fog and edge computing is subtle. Most of the prior literature has treated fog and edge computing as synonymous and has used the word fog and edge computing interchangeably. We clarify the similarities and differences between fog and edge computing. The term \emph{mobile edge computing} (MEC) is also often used in jargon. We point out that MEC is an instance of edge computing where the objective is to provide cloud computing capabilities at the edge of the cellular network. The edge server in MEC is located at the cellular base station. Both fog and edge computing pushes applications, data, services, and computing power away from the centralized nodes to the logical extremes of a network. However, fog computing paradigm has a more \emph{decentralized and distributed control} as compared to edge computing paradigm that has a relatively more centralized control. Another distinction between edge and fog computing is fog's \emph{openness}, which is critical for the success of a ubiquitous fog computing ecosystem for IoT platforms and applications. Proprietary or single vendor solutions, as pursued typically in edge computing, can engender limited supplier diversity, which can have a negative impact on system cost, quality, market adoption, and innovation. Furthermore, \emph{radio access network} in edge computing paradigm is typically a cellular network whereas in fog computing radio access network can be WLAN, WiMax, and/or cellular, and is partially considered a part of the fog.
%
%================================================================%
%                        S. RELATED WORK
%================================================================%
%
\vspace{-2mm}
\section{Related Work}
\label{section_related_work}
Fog computing has been the subject of many research works in recent years. Various fog computing architectures have been proposed \cite{PatelMECWP2014}\cite{maFCMDCDREPM15}\cite{Bittencourt3PGCIC2014}, each of which addresses specific mobility, resource management and optimization issues, however, a universally accepted fog computing architecture and standard has yet to be adopted.

Patel et al. \cite{PatelMECWP2014} discussed the key market drivers, benefits, requirements, objectives, and challenges of MEC. The paper also presented a high-level architectural blueprint for MEC. Aazam et al. \cite{maFCMDCDREPM15} proposed a layered architecture for fog computing to address the resource management challenges, such as resource prediction, allocation, and pricing, in fog servers. The authors used a probability-based model that considered the type, traits and characteristics of fog customers to make these resource management decisions. Bittencourt et al. \cite{Bittencourt3PGCIC2014} proposed a layered architecture to facilitate mobility of connected IoT nodes. In their approach, a virtual machine (VM) instance was created for each IoT node connected to the fog server. When an IoT device crossed the radio boundary of the fog server, then it was handed off to another fog server by exchanging snapshot of the IoT device's VM instance. Their layered fog architecture supported the VM migration. Paglierani \cite{ppHPCNFV15} discussed the use of hardware accelerators in network nodes to support fog computing. The paper demonstrated that combination of hardware acceleration and advanced networking concepts, such as SDN and NFV, can significantly improve network performance.

Although prior works have proposed several fog architectures to address specific issues, however, a reconfigurable fog node architecture that is able to adapt according to application requirements has not been studied. In this article, we propose the IFCIoT architecture that unifies IoT, fog, and cloud computing paradigms to help realization of future IoT and CPS applications. We further propose a reconfigurable fog node architecture that analyzes the applications' characteristics and reconfigure the architectural resources to better meet the peak workload demands.
%
%================================================================================================================%
%                          S. Fog Computing for Intelligent Transportation Systems
%================================================================================================================%
%
\vspace{-2mm}
\section{Fog Computing for Intelligent Transportation Systems}
\label{section_fog_computing_for_ITS}

In this section, we discuss fog computing as a key driver for future intelligent transportation system (ITS) implementations. We begin by describing different agents that constitute an ITS. We then present a classification of scenarios for ITS deployment, the shortcomings of modern ITS implementations in these scenarios and how fog computing implementations potentially overcome these shortcomings. Finally, we list out benefits that different ITS agents get from fog computing based ITS implementations.

\begin{figure}[H]
  \centering
  \includegraphics[width = 3.5in, bb = 14 14 716 458] {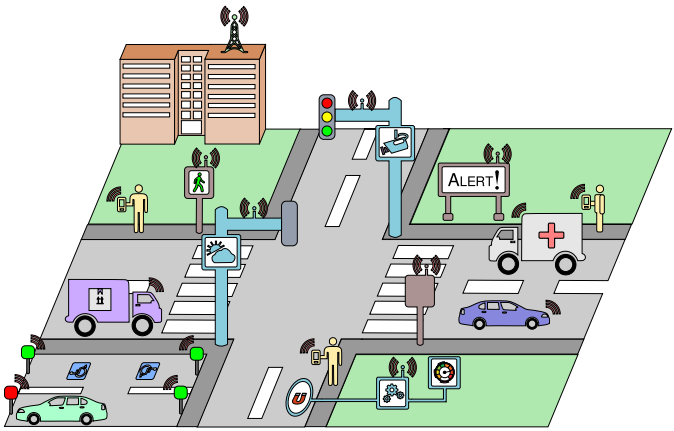}
  \caption[]{Intelligent Transportation System}
  \label{figure_its}
\end{figure}

An ITS consists of different agents such as vehicles, traffic infrastructure and pedestrians (as shown in Fig. \ref{figure_its}). Modern vehicles and traffic infrastructures have a host of integrated electronic subsystems. Vehicles employ electronic subsystems to include features such as, driving assistance, safety and security, infotainment, navigation etc. Traffic infrastructures have sensor systems that sense the number and speed of vehicles through an intersection, road side units for monitoring weather, real-time video surveillance cameras, and, signal and alert systems. Pedestrians also have electronic systems in the form of smart devices which hold information relevant to ITS such as location, direction of walking, walking speed etc. The data generated by each of these ITS agents is vital to implementing an effective ITS. The data has to be processed and communicated to other ITS agents in the system. We discuss data processing and communication mechanisms in modern ITS implementations and compare them with fog computing based ITS implementation for different scenarios in the following text.

An ITS implementation can be broadly classified into two scenarios -- urban and rural. In an urban scenario, there are higher number of agents in the ITS. Providing real-time response to a large number of agents requires highly reliable computation and communication resources. Modern ITS implementations rely on the cloud for computing resources and on cellular networks for communication. Although cellular networks facilitate communication in ITS, they are primarily dedicated to mobile telephony. In an urban scenario, wherein mobile telephony traffic is high, cellular networks (3G and LTE) cannot provide high reliability communication for ITS \cite{HouVFC2016}. Also, the cloud is not a reliable computing resource for ITS. For an effective implementation of a single ITS instance (e.g. at one of the many traffic intersections in a city), data from multiple different agents has to be processed to generate information relevant from each agent's perspective. For a full ITS implementation (e.g. city-wide traffic intersections), an enormous number of unique ITS instances have to be processed. In cloud computing based ITS implementations, all these processing operations are carried out on the cloud. This places a massive burden on the cloud's computing resources. This burden is exacerbated during peak traffic hours, when ITS agents have to be rerouted from traffic congested areas. During such times when the cloud's computing resources are overwhelmed, there is a higher possibility of significant latency in providing response to ITS agents. Late responses during emergency situations like traffic accidents or natural disasters could leed to massive casualties and fatalities.

Fog computing offers higher reliability and flexibility in ITS implementation as compared to using cellular networks and cloud computing. In urban areas, due to large number of ITS agents, movement of traffic is slow. Slow moving traffic opens up the opportunity of using other modes of communication than cellular networks. For example, ITS agents can communicate over a close range using wifi hotspots with multihop communication wherein data is communicated from a source ITS agent to a destination ITS agent by means of intermediate/relay ITS agents. The distributed network of fog computing nodes further facilitates multihop communication. Firstly, the distributed network of fog nodes can be used to provide wifi connectivity to all agents of ITS. Secondly, fog nodes can also serve as intermediate hops in communication. They can perform data filtering and analysis operations on the communicated data to reduce data size. This helps to preserve the bandwidth of the network as well as reduces latency.

Fog nodes also help to unload the processing burden of the cloud in ITS implementation. Instead of having all data processing operations of a full ITS implementation carried out on a centralized cloud, the processing is carried out closer to the edge of the network using a number of distributed fog nodes. Fog nodes process data in a local context i.e. each node processes data for either a single or a small group of ITS instances. This approach can be leveraged because traffic data from an ITS instance is usually only pertinent to that instance or to a few neighboring instances. The fog nodes only communicate summaries of local traffic data to the cloud which uses it for data analytics (traffic patterns, planning construction of new roads etc.). In cases of emergency and disaster situations, local traffic data from all ITS instances have global scope \cite{KhalidEvacSys2016}. In these situations, fog nodes communicate data to the cloud more frequently. Since the fog nodes communicate filtered and locally analyzed data to the cloud, the cloud can swiftly determine an appropriate response.

\begin{figure*}[th]
  \centering
  \includegraphics[width = 6in, bb = 26 54 2351 1625] {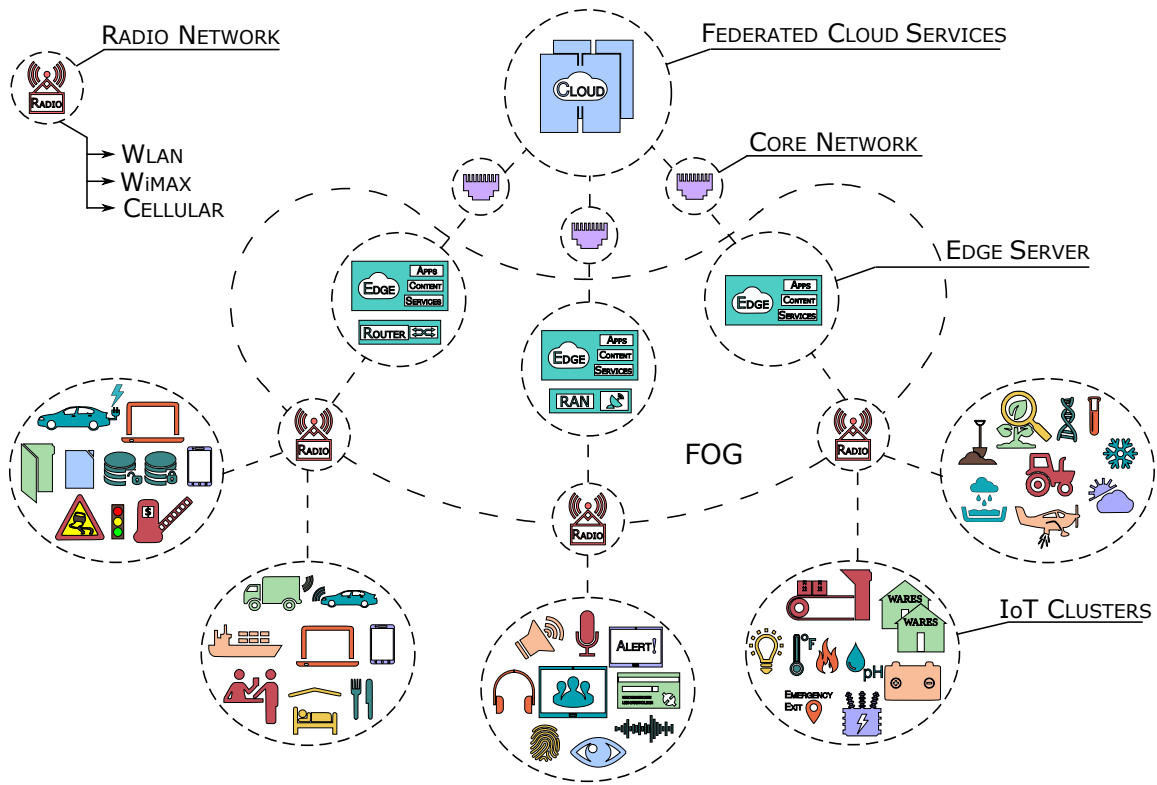}
  \caption[]{IFCIoT architectural paradigm.}
  \label{figure_ifciot_architectural_paradigm}
\end{figure*}

In a rural scenario, there are fewer number of agents in the ITS. Real-time response to a small number of ITS agents can be delivered with significantly less computing and communication resources so, a sparser distribution of fog nodes may be used. In such scenarios, fog nodes are useful in emergency and disaster situations. During mass evacuation of urban areas in disaster situations, traffic may be heavily routed through rural areas. The fog nodes in rural areas help in the management of the increased traffic load by providing reliable computing and communication resources to ITS agents.

Fog computing benefits all ITS agents by improving ITS services. For vehicles, fog computing can provide the following improved services: rerouting from heavy traffic areas (during peak hours), repair or towing services, services in case of accidents, emergency evacuation routes, finding parking space etc. Fog computing helps in development of traffic infrastucture. Traffic flow data collected from fog nodes can be used for the following: changing location of signs and signals based on traffic data analysis, surveying road surfaces for damage, planning construction of new roads etc. Pedestrians have more safety in crosswalks on busy streets. They are also informed about shorter routes for walking to their destinations. Fog computing also improves transit services with features like accurate arrival/departure times, delay and cancellation notifications, passenger count, passengers per stations, vehicle operation duration for repair and maintenance, ticket purchase, seat selection, remote check-ins, and information on hotels and restaurants nearby transit stations, etc.

%
%================================================================%
%                          S. IFCIoT
%================================================================%
%
\vspace{-2mm}
\section{IFCIoT: Integrated Fog Cloud IoT Architectural Paradigm}
\label{section_ifciot_architectural_paradigm}
We propose the IFCIoT architectural paradigm as depicted in Fig.~\ref{figure_ifciot_architectural_paradigm}. The novel aspect of this architecture is that the architecture furnishes federated cloud services to IoT devices via intermediary fog. The federated cloud services are provided by a federated cloud that can comprise of multiple internal and external cloud servers to match business and application needs. As shown in Fig.~\ref{figure_ifciot_architectural_paradigm}, the \emph{fog} comprises of fog nodes (e.g., edge servers, smart routers, base stations, gateway devices) and partially radio access networks. In a fog computing environment, much of the processing takes place on a fog node. In the IFCIoT architecture, the entire fog deployment can be located locally (e.g., in case of building automation, a company that manages a single office complex) or the fog deployments can be distributed at local or regional levels that feed information to a centralized parent system and services (e.g., in case of building automation, a large commercial property management company). In the IFCIoT architecture, each operational fog node is autonomous to ensure uninterrupted operations of the facility/service it provides.

A fog node in the IFCIoT architecture manages all IoT devices that are within its radio network. The IoT devices typically leverage radio access networks (e.g., WLAN, WiMAX, cellular networks) to communicate with the fog whereas the fog is connected to the federated cloud servers via core network. A fog can be connected to other fogs through a radio access network. Specifically, when an IoT device moves from the coverage of one fog to another, the virtual machines associated with the IoT device are migrated from the original host edge server to the migrated edge server \cite{Bittencourt3PGCIC2014}. The fog nodes in the IFCIoT architecture facilitates the collection and maintenance of local system statistics and/or locally sensed information supplied by various IoT devices and/or clusters. These local statistics and information can either be used to improve the local content, services, and applications or to update the federated cloud data center. The federated cloud data center receives updates from multiple fog nodes. The federated cloud data center can then perform big data analytics on the received information to extract information that is representative of a bigger geographical location and to determine global system statistics.
%
%================================================================%
%                        S. FOG ARCHITECTURE
%================================================================%
%
\vspace{-2mm}
\section{Fog Architecture}
\label{section_fog_architecture}
The fog comprises of fog nodes and partially radio access networks as depicted in Fig.~\ref{figure_ifciot_architectural_paradigm}.
This section discusses the radio access network and our proposed fog node architecture for the IFCIoT architectural paradigm.
%
%================================================================%
%                     SS. RADIO ACCESS NETWORK
%================================================================%
%
\subsection{Radio Access Network}
\label{subsection_radio_access_network}
IoT end devices can leverage a multitude of wireless access technologies, such as WLAN, WiMAX, and cellular access networks (e.g., 4G, 5G), as the radio access network for accessing the fog. According to the OpenFog Consortium, fog nodes are not completely fixed to the edge, but should be seen as a fluid system of connectivity. Hence, the radio access network can be considered partially a part of the fog architecture. Fog computing enables the design of an energy- and spectral-efficient radio access network, which can be named as fog computing-based radio access network (F-RAN) \cite{PengFRAN2015}. The F-RAN can take advantage of local radio signal processing, cooperative radio resource management, and distributed storage capability of fog nodes to decrease the load on fronthaul (connection between centralized baseband controllers and remote radio heads at cell sites in a new radio access network architecture) and avoid large-scale radio signal processing in the centralized baseband controllers.
%
%=====================================================================================================================%
%                     SS. RECONFIGURABLE AND ADAPTIVE FOG NODE/EDGE SERVER ARCHITECTURE
%=====================================================================================================================%
%
\subsection{Reconfigurable and Adaptive Fog Node/Edge Server Architecture}
\label{subsection_reconfigurable_and_adaptive_fog_node_edge_server_architecture}
Workload analytics on a server/cloud reveals that different applications have different peak load hours at different times \cite{Singh:2010}. We exploit this time-variance of applications' peak workloads to propose a reconfigurable and adaptive multicore architecture for the edge server that can adapt according to the application load being run at a given time to better sustain the projected data velocity, data volume, and real-time requirements of IoT/CPS applications. Our proposed edge server architecture consists of several layers: application layer, analytics layer, virtualization layer, reconfiguration layer, and hardware layer, as shown in Fig.~\ref{figure_fog_architecture}.

\noindent\textbf{Application Layer:} This is the top-most layer of the edge server architecture. This layer consists of application platform services that the edge server can provide to various applications hosted on the edge server. The application platform services provided by the edge server include services for computation offloading, content aggregation, databases and backup, and network information, etc. When an IoT device connected to the edge server requests for a particular application to be executed, a VM environment is created for the application. This means that each application has its own instantiation of VM environment running on the application layer. The application layer for the edge server acts as a PaaS provider, that is, the application layer abstracts the entire edge server architecture to provide a standard platform for the IoT application developers.

\begin{wrapfigure}{r}{0.25\textwidth}
    \includegraphics[width=0.25\textwidth, bb = 8 5 712 1794] {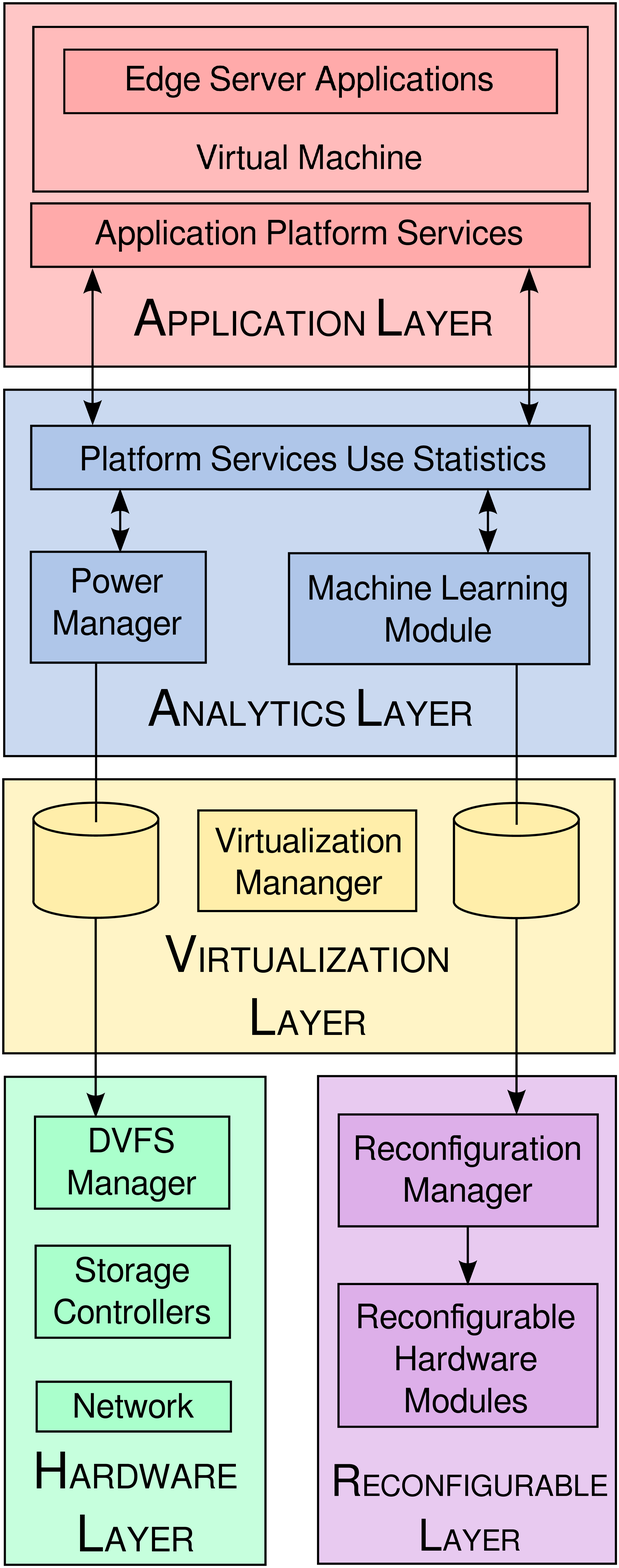}
    \caption{Layered fog node architecture.}
  \label{figure_fog_architecture}
\end{wrapfigure}

\noindent\textbf{Analytics Layer:} The analytics layer consists of three modules: platform services use statistics module, machine learning module, and power manager module. The platform services use statistics module analyzes usage of the application services provided in the application layer. The machine learning module takes the service requests' type and volume information as input and analyzes this information to predict hardware resource requirements, which can be leveraged by the reconfigurable layer. The power manager module analyzes the service requests' type and volume information to determine the edge server utilization. Based on the edge server utilization, the power manager module dynamically adjusts the operating voltage and frequency of the edge server's hardware components.

\noindent\textbf{Virtualization Layer:} The virtualization layer abstracts the underlying hardware resources (hardware resources can be from different vendors) to provide a common interface for application services. The virtualization layer acts as an IaaS provider, that is, the virtualization layer abstracts the hardware resources of the edge server from the application services layer, and hence from the applications running on the edge server.

\noindent\textbf{Reconfigurable Layer:} The reconfigurable layer consists of a reconfiguration manager and a set of reconfigurable modules. The reconfiguration manager in the reconfigurable layer takes input from the machine learning module in the analytics layer, and reconfigures the architectural resources to better meet the requirements of the peak workload application at a given time. The ability to adapt edge server hardware to changing workload requirements is a novel concept which, to our knowledge, has not been addressed in other works \cite{PatelMECWP2014}\cite{Bittencourt3PGCIC2014} in literature.

\noindent\textbf{Hardware Layer:} The hardware layer consists of a dynamic voltage and frequency scaling (DVFS) manager, storage controllers, and network resources. The DVFS module acquires input from the power manager module in the analytics layer. The DVFS module adjusts the operating voltage and frequency of various hardware components of the edge server (e.g., processor core, memory, and peripherals) depending on the workload demands. The storage controllers and storage units are used for database services and backup services. The network module manages the connectivity between the edge server and the IoT devices, and between the edge server and the cloud.
%
%=====================================================================================================================%
%                     S. RECONFIGURABLE AND ADAPTIVE FOG NODE/EDGE SERVER ARCHITECTURE APPLIED TO ITS
%=====================================================================================================================%
%
\vspace{-2mm}
\section{Reconfigurable and Adaptive Fog Node/Edge Server Architecture applied to Intelligent Transportation Systems}
\label{section_reconfigurable_and_adaptive_fog_node_edge_server_architecture_applied_to_its}
In this section, we describe how our proposed reconfigurable and adaptive fog architecture maps to the intelligent transportation system use case that we presented in Section \ref{section_fog_computing_for_ITS}.

\noindent\textbf{Application Layer:} The application layer provides platform services to various applications hosted on the edge server. For example, in an ITS implementation, consider real-time video surveillance as a platform service. This service can be used by vehicles waiting to turn at an intersection to look for vehicles coming from behind or from the sides; by traffic management department for routing traffic; by public transport management department to track its buses; or by emergency response department to access severity of vehicular accidents to dispatch personnel accordingly. All these applications use the same traffic infrastructure, but, process the collected video data differently. All of these applications are hosted on edge servers. Each application is run on a VM environment which hides its processing operations from all other applications.

\noindent\textbf{Analytics Layer:} The analytics layer analyzes the volume of platform service requests and forwards tuning parameters to hardware and reconfigurable layer. For example, consider the case of traffic congestion at an instance of ITS during peak hours. During this time, there are large number of ITS agents requesting platform services from the edge server. The analytics layer detects the increased volume of request and forwards tuning parameters to the reconfigurable layer to increase the processing capabilities of the edge server by instantiating more hardware modules. With increased processing capabilities, more applications can be launched in the application layer.

\noindent\textbf{Virtualization Layer:} The virtualization layer hides the underlying hardware from the application layer by providing a common interface to all hardware modules. For example, a pedestrian can request for platform service from the edge server over a wifi network and a vehicle can request the same platform service but, over an LTE network. The virtualization layer reformats these requests to remove all hardware and network dependency parameters before forwarding it to the application layer. The reformatting process thus, frees up the application layer from all dependencies.

\noindent\textbf{Reconfigurable Layer:} Reconfigurable layer takes input from the analytics layer and reconfigures the architectural resources of the edge server. This increases the flexibility of the edge server and makes it capable to adjust to different workloads. For an ITS implementation, flexibility is advantageous during traffic congestions, as discussed in the analytics layer. Flexibility is also useful in emergency and disaster situations. For example, in the event of a disaster, mass evacuation of several towns and cities has to be carried out. During this time, the number of vehicles on the road would be enormous. Reconfigurable edge servers can adapt to these increased workloads and provide reliable service.

\noindent\textbf{Hardware Layer:} The hardware layer consists of computation and communication hardware components which runs the edge server applications and, reconfigures hardware modules in the reconfigurable layer. In an ITS implementation, the hardware layer components aggregate data from sensors (induction loop sensors, weather sensors, speed-radar sensors etc.) and manage signal and alert displays (traffic signals, warnings for closed roads, bad weather conditions, ongoing construction etc.)
%
%================================================================%
%         S. INSIGHTS INTO OTHER POTENTIAL CONSUMER ELECTRONICS APPLICATIONS
%================================================================%
%
\vspace{-2mm}
\section{Insights into Other Potential Consumer Electronics Applications}
\label{section_insights_into_other_applications}
This section discusses the potential applications of the IFCIoT architecture in various sectors, such as smart cities, localized weather maps and environmental monitoring and, real-time agricultural data analytics and control.

\noindent\textbf{Smart Cities:} The IFCIoT architectural paradigm can provide a basis architecture for various subsystems (e.g., smart grid, smart buildings, industrial plants, hospitals, schools, and law enforcement) in smart cities. A major challenge in establishing smart cities is the requirement of ubiquitous broadband bandwidth and connectivity availability. While most modern cities have multiple cellular networks that provide adequate coverage, these networks often have capacity and peek bandwidth limits that just meet the needs of their existing subscribers. This limited bandwidth of cellular networks makes the realization of advanced municipal services envisioned in a smart city (e.g., real-time surveillance, public safety, on-time advisories, smart buildings) a challenge. The IFCIoT architecture helps in reducing the load on cellular networks by leveraging local radio access networks, local radio signal processing, and cooperative radio resource management in fog nodes. The conserved bandwidth can then be used for providing smart city services.

\noindent\textbf{Localized Weather Maps and Environmental Monitoring:} Localized weather maps can be an interesting application of the IFCIoT architectural paradigm. Various IoT devices measure temperature, humidity, and atmospheric pressure, and send this information to nearby edge servers. The edge servers process the received information from IoT devices to obtain a more refined and localized weather information for customers as opposed to the weather information available from news outlets for the whole city. The edge servers further update the back-end cloud servers for refined weather information and better weather forecasting. Environmental monitoring is a similar application that can be realized in the IFCIoT architectural paragon. The environmental monitoring system that leverages the IFCIoT architecture can provide more localized (e.g., geographical precision ranging from a zip code to less than a mile) and accurate information regarding air quality, allergens, pollution, and noise in an area.

\noindent\textbf{Real-Time Agricultural Data Analytics and Control:} The IFCIoT architecture can improve agricultural health to ensure people's access to safe, plentiful, and nutritious food by enabling real-time agricultural data analytics and automated control where possible. The IoT devices (mainly sensors) in an agricultural area provide localized information regarding soil moisture, precipitation, rain water, water estimate from melting snow, pollution level, pest level, and types of pests to the nearby edge servers. The edge servers process and analyze the received information from the IoT devices in the agricultural field and then determine an accurate scheduling of water sprinkler systems, fertilizers and pesticides supply in the area to preserve the crop quality. Our proposed edge server with real-time analytics engine can provide real-time control of the IoT actuator devices in each agricultural area as opposed to sending and processing all the sensed information in the cloud. The edge servers also raise triggers and alarms for the respective agricultural authorities. The edge servers further update the cloud data centers with the information periodically so that the cloud data center can perform detailed analytics and make agricultural decisions for larger geographical areas.
%
%================================================================%
%                        S. CONCLUSIONS
%================================================================%
%
\vspace{-2mm}
\section{Conclusions}
\label{section_conclusions}
Fog computing provides various advantages over cloud computing for applications that require faster processing with reduced latency and delay jitter, real-time responsiveness, mobility support, and location-based customization. However, fog computing is not a replacement for cloud computing as cloud computing will still be desirable for high end batch processing jobs that are very frequent in the business and scientific worlds. The synergy of fog and cloud computing will help in realization of future IoT and CPS applications. In this article, we have proposed a fog-centric IFCIoT architecture that promises increased performance, energy efficiency, reduced latency, scalability, and better localized accuracy for IoT and CPS applications. To better meet the performance, energy, and real-time requirements of applications, we have also proposed a reconfigurable fog node architecture that can adapt according to the workload being run at a given time. We also elaborate the potential applications of the proposed IFCIoT architecture, such as smart cities, ITS, localized weather maps and environmental monitoring, and real-time agricultural data analytics and control.
%
%================================================================%
%                        S. REFERENCES
%================================================================%
%
{
\balance
\bibliographystyle{IEEEtran}
\bibliography{IEEEabrv,IFCIoTMagazine}
}
%
%================================================================%
%                           BIOGRAPHY
%================================================================%
%
\begin{IEEEbiographynophoto}{Arslan Munir} is an assistant professor in the Department of Computer Science and Engineering at the University of Nevada, Reno. His research interests include embedded and cyber-physical systems, computer architecture, multicore, parallel computing, fault tolerance, and computer security. Munir has a PhD in electrical and computer engineering from the University of Florida, Gainesville. He's a member of IEEE. Contact him at arslan@unr.edu.
\end{IEEEbiographynophoto}

\begin{IEEEbiographynophoto}{Prasanna Kansakar} is a PhD student in the Department of Computer Science and Engineering at the University of Nevada, Reno. His research interests include embedded and cyber-physical systems, computer architecture, multicore, and computer security. Kansakar has a BE in electronics and communication engineering from the Tribhuvan University, Institute of Engineering, Pulchowk Campus, Nepal. Contact him at pkansakar@nevada.unr.edu.
\end{IEEEbiographynophoto}

\begin{IEEEbiographynophoto}{Samee U. Khan} is an associate professor in the Department of Electrical and Computer Engineering at the North Dakota
State University, Fargo. His research interests include optimization, robustness, and security of cloud, grid, cluster and big data computing, social networks, wired and wireless networks, power systems, smart grids, and optical networks. Khan has a PhD from the University of Texas, Arlington. He is a senior member of IEEE, a fellow of the Institution of Engineering and Technology, and a fellow of the British Computer Society. Contact him at samee.khan@ndsu.edu.
\end{IEEEbiographynophoto}
\fussy
}
\end{document}